\documentclass[Physsubmission, Phys]{SciPost}

\hypersetup{
    colorlinks,
    linkcolor={red!50!black},
    citecolor={blue!50!black},
    urlcolor={blue!80!black}
}

\usepackage[bitstream-charter]{mathdesign}
\def\del{\partial}

\newcommand{\eqn}[1]{Eq.~\eqref{#1}}

\long\def\comment#1{ }

\newcommand{\nn}{\nonumber\\ }

\def\be{\begin{eqnarray*}}
\def\ee{\end{eqnarray*}}
\def\beq{\begin{eqnarray}}
\def\eeq{\end{eqnarray}}

\def\bwt{\begin{widetext}}
\def\ewt{\end{widetext}}
\newcommand{\bea}{\beq \begin{aligned}}
\newcommand{\eea}{\end{aligned}\eeq}

\def\k{{\boldsymbol k}}

\def\bell{{\boldsymbol \ell}}

\def\z{{\boldsymbol z}}

\def\0{{\boldsymbol 0}}

\def\k{{\boldsymbol k}}
\def\x{{\boldsymbol x}}
\def\y{{\boldsymbol y}}

\def\r{{\boldsymbol r}}

\def\A{{\boldsymbol A}}

\def\rme{{\rm e}}
\def\rmd{{\rm d}}

\def\tr{ \text{Tr}}

\def\cU{{\cal U}}

\def\cP{{\cal P}}

\DeclareSymbolFont{usualmathcal}{OMS}{cmsy}{m}{n}
\DeclareSymbolFontAlphabet{\mathcal}{usualmathcal}

\begin{document}

\begin{center}{\Large \textbf{
A novel formulation of the unintegrated gluon distribution for DIS \\
}}\end{center}

\begin{center}
Yacine Mehtar-Tani \textsuperscript{1},
\end{center}

\begin{center}
{{\bf 1} Physics Department and RIKEN BNL Research Center, Brookhaven National Laboratory, Upton, NY 11973, USA.}

* mehtartani@bnl.gov
\end{center}

\begin{center}
\today
\end{center}


\definecolor{palegray}{gray}{0.95}
\begin{center}
\colorbox{palegray}{
  \begin{tabular}{rr}
  \begin{minipage}{0.1\textwidth}
    \includegraphics[width=22mm]{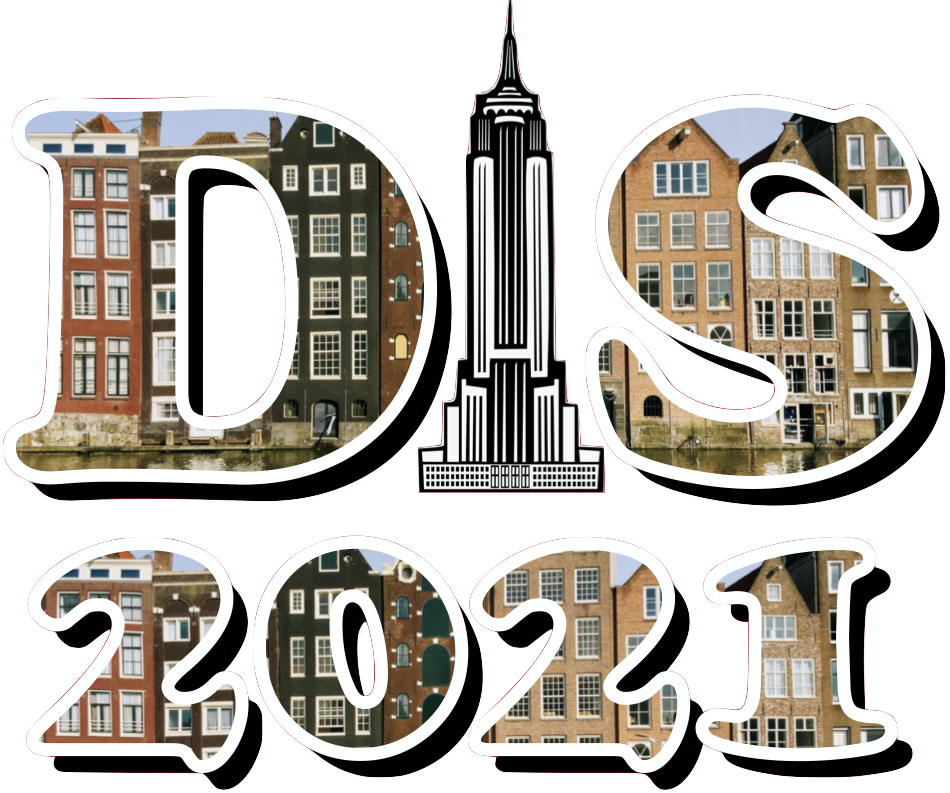}
  \end{minipage}
  &
  \begin{minipage}{0.75\textwidth}
    \begin{center}
    {\it Proceedings for the XXVIII International Workshop\\ on Deep-Inelastic Scattering and
Related Subjects,}\\
    {\it Stony Brook University, New York, USA, 12-16 April 2021} \\
    \doi{10.21468/SciPostPhysProc.?}\\
    \end{center}
  \end{minipage}
\end{tabular}
}
\end{center}

\section*{Abstract}
{\bf
In this talk, we revisit inclusive DIS in the small $x$ limit and derive a new factorization formula that accounts for leading powers in both Bjorken and Regge limits. In this semi-classical description, we obtain a new unintegrated gluon distribution which encompasses both the dipole operator and the gluon Parton Distribution Function with an explicit dependence on the longitudinal momentum fraction $x$. 
}


\section{Introduction}
\label{sec:intro}
In the Bjorken limit of QCD the proton is probed at short distances where it appears to be composed of weakly interacting quarks and gluons owing to the property of asymptotic freedom. The partonic content of the proton is encoded in parton distribution functions (PDF) that measure the probability distribution of partons carrying a longitudinal fraction $x$ of the target momentum at the scale $Q^2$. The latter two variables are related in Deep Inelastic Scattering (DIS) to the virtuality of the exchanged photon in electron-proton collisions and Bjorken $x_{\rm Bj}\sim Q^2/q\cdot P$, where $q$ and $P$ are the virtual photon and proton 4-momenta, respectively. The renormalization group equations that describe the evolution of PDF's as function of virtuality of the process at weak coupling, i.e., the DGLAP equations, resum ladder diagrams along which the transverse momenta of partons are strongly ordered and decrease from $Q^2$ down the non-perturbative scale while the $x$ variable is of order one. 

There is another regime of QCD, the Regge limit, that is achieved at asymptotically high energy $s\sim q\cdot P \gg Q^2$, or equivalently small $x$ providing a large logarithmic phase space for strongly ordered gluons in $k^+= xP^+$ variable.  In contrast with DGLAP evolution, the transverse components of the momenta are typically of the same order along the ladder which implies a strong inverse ordering the $k^-_i\sim k^2_{i \perp}/k^+_i$ components  \footnote{It is customary in collider physics to use light cone variables defined as $k^+=(k_0+k_3)/\sqrt{2}$ and $k^-=(k_0-k_3)/\sqrt{2}$ in addition to the transverse components $\k \equiv (k^1,k^2)$. Hence, in the frame where the proton is moving in the $+z$ direction  its dominant momentum component is $P^+$.}.  At small enough $x$ a remarkable emergent phenomenon is expected to occur: gluon density increases rapidity until it reaches saturation due to non-linear gluon recombination effects as a consequence of unitarity. This takes place at the saturation scale $Q_s$~\cite{glr,mq}, which increases with decreasing $x_{\rm Bj}$. 

Although seemingly different, these two descriptions of hadron structure exhibit similarities at higher orders in the coupling constant. This motivates the search for a unified framework. Furthermore, it was observed recently in Ref.~\cite{negativeXS} that the NLO corrections (see~\cite{nlobfkl,nlobk,nlojimwlk}) to the Balitsky-Fadin-Kuraev-Lipatov (BFKL)~\cite{bfkl}, Balitsky-Kovchegov (BK)~\cite{bkian,bkyuri} and Jalilian-Marian-Iancu-McLerran-Weigert-Leonidov-Kovner (JIMWLK)~\cite{jimwlk} equations that govern small-$x_{\rm Bj}$ physics, yield large collinear logarithms that need to be resummed to ensure numerical stability of the equations~\cite{collinear-logs-Beuf,collinear-logs-Edmond}. This issue if is a consequence of the so-called shock wave approximation which assumes that the target longitudinal extent is equal to zero due to Lorentz contraction of the proton as perceived by the photon and as a result the + and - components of the gluons along the ladder are decoupled and the transverse integrations are not explicitly constrained through the evolution. 

In this talk, I report our attempt to provide a semi-classical approach of DIS at small $x$ beyond the shock wave approximation where the longitudinal extent of the proton is accounted for \cite{Boussarie:2020fpb}.  The leading powers in both the Bjorken and the Regge limits are obtained by performing a gradient expansion around the transverse position of quantum fluctuations. In this approach, light cone time ordering between long-lived quantum fluctuations such as the photon splitting into a $q\bar q$ dipole and the interaction of the dipole with the target field, is always satisfied implying an ordering of both $k^+$ and $k^-$ variables. 

\section{A novel unintegrated gluon distribution }
\begin{figure}
\begin{center}
\includegraphics[width=0.5\textwidth]{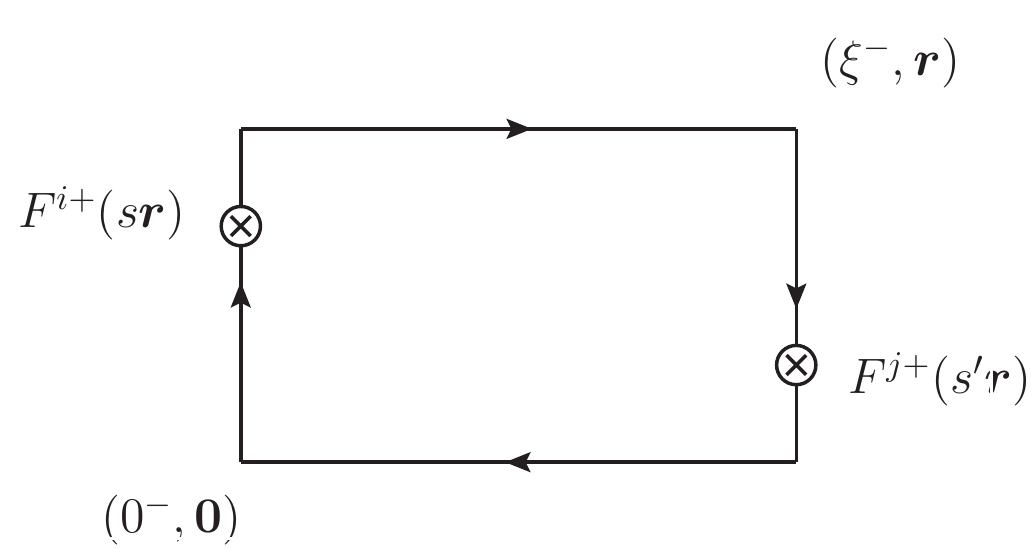} 
\end{center}
\caption{Diagrammatic depiction of the nonlocal operator that defines the unintegrated gluon distribution  in \eqn{eq:dist-def}. The horizontal and vertical lines represent path ordered Wilson lines along the $+$ and transverse directions, respectively.  }
\label{fig:splitting}
\end{figure}

Before we discuss its derivation in the context of inclusive DIS let us first introduction the operator definition of the new unintegrated gluon distribution that corrects for the $x$ dependence of the dipole scattering amplitude, which naturally arises in the small-$x$ limit.  It reads

  \beq\label{eq:dist-def}
&&xG^{ij}(x,\k)  = \int \frac{\rmd \xi^- \rmd^2 \r }{(2\pi)^3 P^+}\, \rme^{i xP^+\xi^-  -i \k\cdot \r} \nn
&&\times\int_0^1 \rmd s \int_0^1 \rmd s'
 \,\langle P|  \, \tr  \,  \cU_\0(s\r,s'\r) F^{i+}(\xi^-,s'\r) \, \cU_\r(s'\r,s\r) F^{j+}(0,s\r) | P \rangle \,,\nn
 \eeq
where  $\k$ is a transverse momentum and $i,j=1,2$ label two orthogonal transverse directions. Furthermore, $F^{i+}=\del^iA^+-\del^+A^i-ig[A^i,A^+]$ is the field strength tensor
and 
 \beq
 \cU_\0(s\r,s'\r)=   [s\r,\0]_{0^-}[0,\xi^-]_\0[\0,s'\r]_{\xi^-}\,,\,\quad \cU_\r(s'\r,s\r) =   [s'\r,\r]_{\xi^-}  [\xi^-,0]_\r[\r,s\r]_{0^-}\,, 
 \eeq
 are two finite length staple-shaped gauge links that connect  $F^{i+}(\xi^-,s'\r)$ to $F^{i+}(0^-,s\r)$ as depicted in Fig.~ \ref{fig:splitting}~\footnote{$F^{i+}(\xi^-,\r)\equiv F^{i+}(\xi^-,\xi^+=0,\r)$}, and where
\beq
 [\xi^-,0^-]_{\r} \equiv  \cP \exp\left[ig \int^{\xi^-}_{ 0^-} \rmd x^-A^+(x^-,\r)\right]
\quad \text{and} \quad 
 [ \x,\y]_{\xi^-} \equiv \cP \exp\left[-ig \int_{ \y}^{ \x} \rmd \z \cdot  \A(\xi^-,\z)\right]\nonumber
\eeq
are path ordered Wilson lines in the $+$ and $\perp$ directions, respectively, with $\z \equiv \z(s) = s \,\x +(1-s)\,\y$. 

This distribution distribution (\ref{eq:dist-def}) encompasses both the gluon PDF at large $x$ and the dipole unintegrated distribution at small $x$.  
Integrating over $\k$ yields a $\delta(\r)$ and we readily recover the gluon PDF
  \beq\label{eq:dist-def-pdf}
&& \int \rmd^2 \k \, x G^{ii}(x,\k)  =xg(x) \equiv \int \! \frac{\rmd \xi^-}{ (2\pi)P^-} \rme^{i xP^+\xi^- }  \,  \,\langle P|  \, \tr  \, [0,\xi^-] F^{i+}(\xi^-) [\xi^-,0] F^{i+}(0) | P \rangle\,,\nn
 \eeq
where the gluon operator is implicitly evaluated at $\r=0$. Let us turn now to the small $x_{\rm Bj}$ limit. In \eqn{eq:dist-def}  we first set $x=0$, and neglecting the transverse gauge links that can be gauged away along with all $A^i$ fields. 
 Then, to make contact with the dipole amplitude in the small $x_{\rm Bj}$ limit one has to contract our distribution with $\r^i \r^j$. This projection is reminiscent of the standard dipole cross-section formula $\sigma(|\boldsymbol{r}|) \propto \boldsymbol{r}^2 xg(x,1/|\boldsymbol{r}|)$~\cite{Kovchegov:2012mbw}. The dipole scattering amplitude is recovered by using the fact that $-\boldsymbol{r}^i \int_0^1 \rmd s \,  F^{i+}(s\r)=  A^+(\r)-A^+(\0)$, then rewrite the $A^+$ terms as ta derivative acting on $\tr [\xi^-, \zeta^-]_\r [\zeta^- , \xi^-]_\0 $ w.r.t. $\xi^-$ and $\zeta^-$, respectively. Finally, upon integration, we finally obtain that $ \int\!{\rm d}^{2}\boldsymbol{k}\,{\rm e}^{i \boldsymbol{k}\cdot\boldsymbol{r}}\left[\boldsymbol{r}^{i}\boldsymbol{r}^{j}xG^{ij}(x,\boldsymbol{k})\right]_{x=0}
 =\frac{2}{\alpha_{s}}\int\frac{{\rm d}^{2}\boldsymbol{b}}{(2\pi)^{2}}{\rm Re}\langle N_{c}-\mathrm{tr}\left(U_{\boldsymbol{b+r}}U_{\boldsymbol{r}}^{\dagger}\right)\rangle$ 
where $U_\r = [+\infty,-\infty]_\r$ and $\langle \rangle \equiv \langle P | | P \rangle/\langle P | P \rangle $. 
Note that  since the longitudinal phase factor is absent the interactions with the target (encoded in the Wilson line) can take place at any light cone time and are therefore not restricted by the kinematics. This results in the aforementioned violation of the $k^+$ ordering in the shock wave approximation. 

%

%
%
%
\section{Beyond high energy factorization}
We turn now to the factorization formula involving the gluon distribution (\ref{eq:dist-def}). 
The dominant process at high energy stems from the splitting of the virtual photon into a pair of eikonal quark anti-quark that propagate in the $-$ direction in the target background field $A^+$. This amounts to assuming that the latter is evaluated on the light cone $x^+=0$. In the shock wave approximation the splitting is assumed to take place way before the scattering occurs. We shall relax this approximation by keeping the sub-eikonal corrections that account for splittings inside the target. 

In order to factorize the hard matrix element from the target operator we extract the first and last interactions with the target (i.e., the first and last $A^+$ insertions) which result in four contributions that can be combined into one by using the following identity 
\beq \label{eq:AAF}
 A^+(\x+\r)-A^+(\x) = -\r^i \int_0^1 \! \rmd s\, F^{i+}(\x + s\r)\,.
\eeq
This transformation allows for a dynamical notion of the perceived extent of the target. 
The last step is somewhat similar to twist expansion in the Bjorken limit. Instead of expanding in inverse powers of $s$ we perform a gradient expansion of the quark and antiquark propagators around their transverse positions \cite{Boussarie:2020fpb,next-to-eikonal} that is consistent for both small and large $x_{\rm Bj}$.  

Our final result is the following factorization formula for the cross-section of the DIS subprocess $\gamma^\ast(q) +{ \rm proton}\,(P)  \to  X$:

 \beq \label{eq:cross-section}
 \sigma_{\lambda}(x_{\rm Bj}, Q^2)&=&4 \alpha_{\rm em}\alpha_s \sum_f q_f^2 \! \int_0^1 \! \frac{\rmd x}{2\pi}\int_0^1 \! \frac{\rmd z}{2\pi} \int \! \rmd^2 \k \, \rmd^2 \bell \, \del^i \phi_\lambda \! \left(\bell+\frac{\k}{2}\right)\del^j \phi_{\lambda}^\ast \! \left(\bell-\frac{\k}{2}\right) \! \nn
 && \times\delta \! \left( x -x_{Bj}-\frac{\bell^2}{2z\bar z q^-P^+}\right) \! \, x  \, G^{ij}(x,\k),
 \eeq
that, interestingly enough, exhibits the same hard factor as in the shock wave approximation. Indeed, $\phi_\lambda$ stands for the Fourier transform of the standard LO photon wave functions. The delta function relates Bjorken $x$ to Feynman x in the gluon distribution as well as the loop variables. In the strict $s \to \infty$, the $x$  integration is trivial, that is $x=0$. 
Note the derivatives acting of the wave functions, which would yield powers of the dipole size $\boldsymbol{r}^i\boldsymbol{r}^j$ in the small $x_{\rm Bj}$ limit, see Eq.~(\ref{eq:dipole-limit}). 

This equation is exact up to corrections of relative order $p_\perp/\sqrt{s}$, where $p_\perp$ is an intrinsic transverse momentum in the proton. Such corrections are suppressed in the Bjorken regime as well as in the Regge limit. The explicit $x$ dependence in 
Eqs.~(\ref{eq:cross-section}) results in a non-locality in transverse dipole sizes which is was shown to be not compatible with the dipole model in \cite{Bialas:2000xs}.

\section{Conclusion}
To summarize, we have generalized high energy factorization approach beyond the shock wave approximation by properly treating the longitudinal structure of the target that restores the explicit $x$ dependence of the gluonic operator lost in strict Regge limit at LO and thus, accounts systematically for both $k^+$ and $k^-$ ordering. Furthermore, this new approach yields a new unintegrated gluon distribution that interpolates between the Regge and Bjorken limits whose quantum evolution is left for future work. 

\section*{Acknowledgements}


\paragraph{Funding information}
This work was supported by the U.S. Department of Energy, Office of Science, Office of Nuclear Physics, under contract No. DE- SC0012704. We acknowledge support from the RHIC Physics Fellow Program of the RIKEN BNL Research Center. 


%
%

\bibliographystyle{apsrev4-2}






\end{document}